\documentclass[preprint, superscriptaddress]{revtex4}
\usepackage{graphicx}
\usepackage{epstopdf}
\usepackage{amsmath, amsfonts, amssymb, bm}
\begin{document}
\title{Entanglement of a laser driven pair of two-level qubits via its phonon environment}
\author{Elena \surname{Cecoi}}
\affiliation{Institute of Applied Physics, Academy of Sciences of Moldova, Academiei str. 5, MD-2028 Chi\c{s}in\u{a}u, Moldova}
\author{Viorel \surname{Ciornea}}
\affiliation{Institute of Applied Physics, Academy of Sciences of Moldova,
Academiei str. 5, MD-2028 Chi\c{s}in\u{a}u, Moldova}
\author{Aurelian \surname{Isar}}
\affiliation{Horia Hulubei National Institute of Physics and Nuclear Engineering, 
Reactorului str. 30, P.O.BOX MG-6 Bucharest - Magurele, Romania}
\author{Mihai  A. \surname{Macovei}}
\affiliation{Institute of Applied Physics, Academy of Sciences of Moldova,
Academiei str. 5, MD-2028 Chi\c{s}in\u{a}u, Moldova}
\date{\today}
\begin{abstract}
The entanglement dynamics of a laser-pumped two-level quantum dot pair is investigated in the steady-state. The closely spaced two-level emitters, 
embedded in a semiconductor substrate, interact with both the environmental vacuum modes of the electromagnetic field reservoir as well as with 
the lattice vibrational phonon thermostat.  We have found that the entanglement among the pair's components is substantially enhanced due to 
presence of the phonon subsystem. The reason is phonon induced decay among the symmetrical and antisymmetrical two-qubit collective states and, 
consequently, the population of the latter one. This also means that through thermal phonon bath engineering one can access the subradiant 
two-particle cooperative state.
\end{abstract}
\maketitle
\section{Introduction}
Quantum dot qubits systems have been studied extensively in the last years. A substantial experimental progress involving these systems has been 
made. The interest in quantum dots lies in the fact that their properties can be engineered and to some degree they behave as real atoms. From 
these reasons often quantum dots systems, together with quantum wells or quantum circuits samples, are called artificial atomic systems and are 
relevant constituents towards quantum technologies. As a consequence, entangled photons generation via high-symmetry site-controlled quantum 
dots was already demonstrated \cite{expp}, whereas single- or cascaded-photon emission from the Mollow triplet sidebands was experimentally 
proven \cite{exp} as well as sub-natural single-photon emissions from a single quantum dot \cite{sub}. In certain laser-pumped artificial atomic samples, 
phonon environment may lead to broadening of spectral lines or to asymmetrical peaks in the Mollow spectrum \cite{assm}. Furthermore, 
temperature-dependent Mollow triplet spectra from a single quantum dot was observed as well \cite{exp2}, while self-homodyne measurement of a 
dynamic Mollow triplet in the solid state was reported too \cite{exp3}. Notice earlier works on control of the quantum dynamics in these systems, see 
for instance \cite{chk,psp}. Moreover, few- or many-qubit ensembles are as well relevant for present or future quantum applications. Particularly, the 
interaction among closely spaced artificial emitters are dominated by dipole-dipole coupling and by phonon-assisted energy transfer processes \cite{exp4}. 
The fluorescence properties of nearly identical quantum dots systems were largely investigated \cite{fluor}, including phonon spontaneous emission 
\cite{sp_pn}. Photon superradiance phenomenon in quantum dot samples was observed experimentally \cite{exp1} whereas fast phonon dynamics in 
optomechanical systems, due to cooperative effects among closely placed multiple two-level quantum dots, was theoretically reported as well \cite{victor}. 

Thus, a quantum dot ensemble may mutually couple via interaction with photon or phonon subsystems, respectively. Therefore, one may naturally 
ask about the entanglement creation among the quantum dots. In this context, entanglement between two quantum dots in a cavity injected 
with squeezed vacuum was demonstrated in Ref.~\cite{gao}. Entanglement dynamics between two coupled quantum dots in a nanowire 
photonic-crystal system was investigated in \cite{ent1}. Recently, it was demonstrated that the degree of entanglement can be controllably 
tuned during the time evolution of quantum dots system \cite{ent2}. Notably, phonon-mediated generation of quantum correlations between 
quantum dot qubits was proven to be efficient in a certain temperature range, but not at lower ones \cite{corrs}. Additionally, complete 
disentanglement by phonon induced dephasing of a pair of two-level qubits was theoretically predicted in Ref.~\cite{def}, see also \cite{deff}.

In contrast, here, we shall demonstrate the steady-state entanglement among a laser-pumped pair of two-level closely packed quantum dots 
where the phonon environmental thermostat significantly enhances this particular quantum effect at lower temperatures. The reason consists 
in the fact that the phonon environmental reservoir opens a decay channel between the symmetrical and antisymmetrical two-qubit collective 
states, respectively. At lower temperatures this allows to populate the subradiant state depending on the strength and sign of the dipole-dipole 
interaction potential. However, higher phonon bath temperatures do not lead to substantial entanglement creation among the quantum dots as 
long as the laser-qubit coupling strength, i.e. Rabi frequency, is of the order of the dipole-dipole frequency shift between the superradiant and 
subradiant two-particle collective states, or less.

The article is organized as follows. In Sec. II we describe the analytical approach and the system of interest, whereas in Sec. III we analyze the 
obtained results. A summary is given in Sec. IV.

\section{Theoretical framework}
We consider a laser pumped system consisting of a pair of identical two-level quantum dots. The laser wave-vector is perpendicular to the line connecting 
the qubits. The quantum emitters interact also with the environmental vacuum modes of the electromagnetic field reservoir as well as with the vibrational 
phonon thermostat. 

In the following, we shall present the corresponding master equation describing the investigated model where the vacuum modes of the electromagnetic field 
reservoir as well as the phonon thermostat are included properly.
\subsection{The master equation}
The master equation describing this system in the absence of phonons and in the Born-Markov approximations, is given as follows \cite{leh,leh1,fik,noi}
\begin{eqnarray}
\dot \rho &+&\frac{i}{\hbar}[H,\rho]=-\frac{\gamma}{2}(1+\chi_{r})[R_{es}+R_{sg},(R_{se}+R_{gs})\rho] \nonumber \\
&-&\frac{\gamma}{2}(1-\chi_{r})[R_{ea}-R_{ag},(R_{ae}-R_{ga})\rho] + H.c.,
\label{ros}
\end{eqnarray}
where an overdot denotes differentiation with respect to time. In Eq.~(\ref{ros}), the Hamiltonian characterizing the coherent quantum dynamics of the 
qubits interacting with the laser field is $H=H_{0}+H_{i}$, where
\begin{eqnarray}
H_{0}/\hbar = 2\Delta R_{ee} + (\Delta+\Omega_{dd})R_{ss} + (\Delta - \Omega_{dd})R_{aa},
\label{h0}
\end{eqnarray}
and
\begin{eqnarray}
H_{i}= \sqrt{2}\hbar\Omega(R_{es}+R_{sg} + R_{se}+R_{gs}). \label{hi}
\end{eqnarray}
Here, $\Omega$ denotes the standard Rabi frequency, whereas $\Delta$ is the detuning of the two-level qubit's frequency from the laser one. 
$\chi_{r}$ describes the radiative coupling among the two-level qubits while $\Omega_{dd}$ corresponds to the dipole-dipole interaction potential, 
respectively. The radiative coupling $\chi_{r}$ goes to zero (unity) for larger (smaller) interparticle separations $r$ in comparison to the photon 
emission wavelength. Correspondingly, $\Omega_{dd}$ tends to zero or to the static dipole-dipole interaction potential, namely, $\Omega_{dd}=
3\gamma(1-3\cos^{2}{\zeta})/4(kr)^{3}$, where $\zeta$ is the angle between the transition dipole vector $\vec d$ and the vector connecting the two 
qubits, i.e., $\vec r$. The single-qubit spontaneous decay rate is $\gamma=k^{3}d^{2}/6\pi\epsilon\epsilon_{0}\hbar$, where $k$ is the resonant 
photon wave-vector in the medium, while $\epsilon_{0}$ is the vacuum permitivity whereas $\epsilon$ is the relative dielectric constant of the semiconductor.
The two-qubit transition operators are obtained using the common Dicke states, namely, $R_{\alpha\beta}$=$|\alpha\rangle\langle \beta|$, 
where $\{\alpha,\beta\} \in \{e,g,s,a\}$ denote the two-qubit excited state and the ground state, and the symmetrical and antisymmetrical collective 
states, respectively. They obey the commutation relation: $[R_{\alpha\beta},R_{\beta'\alpha'}]$ = $\delta_{\beta\beta'}R_{\alpha\alpha'}$ - 
$\delta_{\alpha'\alpha}R_{\beta'\beta}$. Notice that the collective two-qubit states are defined as follows \cite{leh,leh1,fik,noi}
\begin{eqnarray}
|e\rangle = |ee\rangle, ~~~|s\rangle = \frac{1}{\sqrt{2}}\bigl(|eg\rangle + |ge\rangle\bigr),   \nonumber \\
|a\rangle = \frac{1}{\sqrt{2}}\bigl(|eg\rangle - |ge\rangle\bigr), ~~~|g\rangle = |gg\rangle.
\label{dks}
\end{eqnarray}
The first term describing the damping phenomenon in the master equation (\ref{ros}) is responsible for the spontaneous decay via the 
symmetrical channel $|e\rangle \to |s\rangle \to |g\rangle$, whereas  the second one through the antisymmetrical channel 
$|e\rangle \to |a\rangle \to |g\rangle$, respectively. The antisymmetrical transitions are less available when the two qubits get closer to each other 
because $\gamma(1-\chi_{r})/2  \ll \gamma$ in this particular case. As we shall demonstrate further, things change in the presence of phonons.

In the presence of vibrational phonon reservoir one has to modify the master equation accordingly, that is, we have to add the corresponding 
contribution \cite{defff,pn1,pn2,pn3}. The Hamiltonian, $H_{pn}$,  describing the interaction of the qubit subsystem with the phonon reservoir can be 
represented as
\begin{widetext}
\begin{eqnarray}
H_{pn} &=& \sum_{p}\hbar\omega_{p}b^{\dagger}_{p}b_{p} + H_{0} + \frac{i}{2}\sum_{p1}\lambda^{(1)}_{p1}
\bigl(2R_{ee}+R_{aa}+R_{ss}+R_{sa}+R_{as}\bigr)
\bigl(b^{\dagger}_{p1}-b_{p1}\bigr) \nonumber \\
&+& \frac{i}{2}\sum_{p2}\lambda^{(2)}_{p2}\bigl(2R_{ee}+R_{aa}+R_{ss}-R_{sa}-R_{as}\bigr)\bigl(b^{\dagger}_{p2}-b_{p2}\bigr).
\end{eqnarray}
\end{widetext}
Eliminating the phonon bosonic operators $\{b^{\dagger}_{p},b_{p}\}$ in a standard way, i.e., by assuming weak qubit-phonon coupling strengths 
as well as neglecting memory effects, we arrive at a master equation describing the qubit subsystem interaction with the phonon thermostat only. 
The corresponding phonon decay rates are calculated to second order in the coupling strengths $\lambda^{(\xi)}_{p\xi}$, $\{\xi \in 1,2\}$. Then, 
taking into account the contribution arising from the photon reservoir, that is Eq.~(\ref{ros}), the final master equation describing the whole system 
is:
\begin{eqnarray}
\dot \rho &+& \frac{i}{\hbar}[H,\rho]=-\frac{\gamma}{2}(1+\chi_{r})[R_{es}+R_{sg},(R_{se}+R_{gs})\rho] \nonumber \\
&-&\frac{\gamma}{2}(1-\chi_{r})[R_{ea}-R_{ag},(R_{ae}-R_{ga})\rho] \nonumber \\
&-&\Gamma(1+\bar n)[R_{sa},R_{as}\rho] - \Gamma\bar n[R_{as},R_{sa}\rho] + H.c..
\label{rof}
\end{eqnarray}
Here $\Gamma=\frac{\pi}{4}\sum_{\xi=\{1,2\}}\sum_{p\xi}(\lambda^{(\xi)}_{p\xi}/\hbar)^{2}\delta(\omega_{p\xi}-2\Omega_{dd})$, with $\Omega_{dd} > 0$, 
is the decay rate among the symmetrical and antisymmetrical two-qubit states due to phonon thermostat. Particularly, this decay 
rate can be approximately represented as \cite{exp_rr}: $\Gamma \approx (A\hbar/\pi k_{B})(2\Omega_{dd})^{3}\exp{[-(2\Omega_{dd}/\omega_{c})^{2}]}$, 
with $\omega_{c}$ being the cutoff frequency. For typical parameters $A \sim 11{\rm fs/K}$, $\omega_{c} \sim 3\times10^{12}{\rm Hz}$,  and 
$2\Omega_{dd} \sim 10^{12}{\rm Hz}$ one has $\Gamma \sim 10^{10}{\rm Hz}$. Further, the interparticle separation was assumed to be 
larger than the main phonon bath wavelength as well as the linear dimensions of a single quantum dot. Respectively, it is considered to be smaller than 
the relevant photon wavelength. Hence, the two-level emitters interact with independent phonon reservoirs. Respectively, the average mean phonon 
number $\bar n$, at temperature $T$, is given by the expression $\bar n=[\exp{(2\hbar\Omega_{dd}/k_{B}T)}-1]^{-1}$, with $k_{B}$ being the 
Boltzmann constant. If $\Omega_{dd} < 0$, that is $\Omega_{dd} = - |\Omega_{dd}| \equiv -\Omega_{dd}$, the corresponding Master Equation can 
be obtained from Eq.~(\ref{rof}), making the transformation $s \leftrightarrow a$  in Eq.~(\ref{h0}) and in the last line of (\ref{rof}). Thus, generalizing 
at this stage, one can observe a phonon induced decay rate between the collective states $|s\rangle \leftrightarrow |a\rangle$ around the dipole-dipole 
frequency shift between these states. Finally, the scheme works also if one considers an identical dipole-dipole coupled qubit pair consisting from a two-level 
atom and a two-level quantum dot, respectively, i.e. in this case either $\lambda^{(1)}_{p1}$ or $\lambda^{(2)}_{p2}$ is zero.

\subsection{The equations of motion}
The equations of motion describing the quantum dynamics of a pair of laser-pumped two-level quantum dots interacting also with the environmental vacuum 
and thermal phonon reservoirs can be easily obtained from the master equation (\ref{rof}). For our purpose,  when $\Omega_{dd}>0$, the equations of 
motion for the populations of the two-qubit collective states are, respectively,
\begin{eqnarray}
\langle \dot R_{ee}\rangle &=& i\sqrt{2}\Omega\bigl(\langle R_{se}\rangle - \langle R_{es}\rangle\bigr) - 2\gamma\langle R_{ee}\rangle, \nonumber \\
\langle \dot R_{ss}\rangle &=& i\sqrt{2}\Omega\bigl(\langle R_{es}\rangle - \langle R_{se}\rangle + \langle R_{gs}\rangle - \langle R_{sg}\rangle\bigr) \nonumber \\
&-&\bigl(\gamma(1+\chi_{r})+2\Gamma(1+\bar n)\bigr)\langle R_{ss}\rangle +\gamma(1+\chi_{r})\langle R_{ee}\rangle \nonumber \\
&+& 2\Gamma\bar n \langle R_{aa}\rangle, \nonumber \\
\langle \dot R_{aa}\rangle &=& \gamma(1-\chi_{r})\langle R_{ee}\rangle - \bigl(\gamma(1-\chi_{r})+2\Gamma\bar n\bigr)\langle R_{aa}\rangle \nonumber \\
&+& 2\Gamma(1+\bar n)\langle R_{ss}\rangle,
\label{sep}
\end{eqnarray}
with 
\begin{eqnarray*}
\langle R_{ee}\rangle + \langle R_{ss}\rangle + \langle R_{aa}\rangle + \langle R_{gg}\rangle =1. 
\end{eqnarray*}
This system of equations is not closed and, therefore, we present the equations of motion for the two-qubit coherences, namely,
\begin{eqnarray}
\langle \dot R_{es}\rangle &=& i\bigl(\Delta - \Omega_{dd} + i\gamma(3+\chi_{r})/2 +i\Gamma(1+\bar n)\bigr) \langle  R_{es}\rangle \nonumber \\
&+&i\sqrt{2}\Omega\bigl(\langle R_{ss}\rangle -  \langle R_{ee}\rangle -\langle R_{eg}\rangle\bigr), \nonumber \\
\langle \dot R_{sg}\rangle &=& i\bigl(\Delta + \Omega_{dd} + i\gamma(1+\chi_{r})/2 + i\Gamma(1+\bar n)\bigr)\langle  R_{sg}\rangle \nonumber \\
&+&i\sqrt{2}\Omega\bigl(\langle R_{eg}\rangle +  \langle R_{gg}\rangle -\langle R_{ss}\rangle\bigr) + \gamma(1+\chi_{r})\langle  R_{es}\rangle, \nonumber \\
\langle \dot R_{eg}\rangle &=&\bigl(2i\Delta - \gamma)\langle  R_{eg}\rangle +i\sqrt{2}\Omega\bigl(\langle R_{sg}\rangle -  \langle R_{es}\rangle \bigr).
\label{sec}
\end{eqnarray}
The missing equations of motion can be obtained via Hermitian conjugation of the above system of three equations, i.e., Eqs.~(\ref{sec}).
One can observe that the system of equations (\ref{sep}) and (\ref{sec}) involve mostly the transitions 
$|e\rangle \leftrightarrow |s\rangle \leftrightarrow |g\rangle$. Importantly, there is a phonon induced population decay in the antisymmetrical two-qubit 
collective state through the channel $|s\rangle \to |a\rangle$ when $\Omega_{dd} > 0$, see the last equation in (\ref{sep}). This would change the qubits 
quantum dynamics in an external laser field and subject to damping via photon and phonon surrounding reservoirs. 

In the absence of a cw coherent driving, i.e. when $\Omega=0$, the equations of motion Eq.~(\ref{sep}) and Eq.~(\ref{sec}) can be solved analytically. 
For instance, the populations of the cooperative two-qubit states are given by the following expressions:   
\begin{eqnarray}
&{}&\langle R_{ss}(t)\rangle =\frac{e^{-\Gamma_{+}t}}{\bar \Omega}\biggl(\sinh{(\bar\Omega t)}\bigl(2\bar n\Gamma R_{aa}(0)-(\Gamma+\gamma\chi_{r}) 
\nonumber \\
&\times& R_{ss}(0)\bigr) + \bar\Omega R_{ss}(0)\cosh{(\bar\Omega t)}\biggr) + \frac{\gamma R_{ee}(0)e^{-\Gamma_{+}t}}{\bar\Omega
(\Gamma^{2}_{-}-\bar\Omega^{2})}\nonumber \\
&\times& \biggl(\bar\Omega\cosh{(\bar\Omega t)}\bigl(\gamma(1+\chi_{r})^{2}-4\bar n\Gamma\bigr) + \bar \alpha\sinh{(\bar\Omega t)} \biggr) \nonumber \\
&+& \frac{\gamma R_{ee}(0)e^{-2\gamma t}}{\Gamma^{2}_{-}-\bar\Omega^{2}}\biggl(4\bar n\Gamma - \gamma(1+\chi_{r})^{2}\biggr), \nonumber \\
&{}&\langle R_{aa}(t)\rangle = \frac{e^{-\Gamma_{+}t}}{\bar \Omega}\biggl(\sinh{(\bar\Omega t)}\bigl(2\Gamma(1+\bar n)R_{ss}(0) \nonumber \\
&+& (\Gamma+\gamma\chi_{r})R_{aa}(0)\bigr) + \bar\Omega R_{aa}(0)\cosh{(\bar\Omega t)}\biggr) \nonumber \\
&+&\frac{\gamma R_{ee}(0)e^{-\Gamma_{+}t}}{\bar\Omega(\Gamma^{2}_{-} - \bar\Omega^{2})}\biggl(\bar\Omega\cosh{(\bar\Omega t)}\bigl(\gamma(1-\chi_{r})^{2}-4\Gamma(1+\bar n)\bigr) \nonumber \\
&+& \bar\beta\sinh{(\bar\Omega t)} \biggr) + \frac{\gamma R_{ee}(0)e^{-2\gamma t}}{\Gamma^{2}_{-}-\bar\Omega^{2}}\biggl(4\Gamma(1+\bar n) 
\nonumber \\
&-& \gamma(1-\chi_{r})^{2}\biggr), \nonumber \\
&{}&\langle R_{ee}(t)\rangle=\langle R_{ee}(0)\rangle e^{-2\gamma t}, \nonumber \\
&{}&\langle R_{gg}(t)\rangle = 1 - \langle R_{ee}(t)\rangle -\langle R_{ss}(t)\rangle - \langle R_{aa}(t)\rangle,
\label{soll}
\end{eqnarray}
where $\bar \alpha=\Gamma_{-}\bigl\{\gamma\chi_{r}(1+\chi_{r})+\Gamma\bigl(1+\chi_{r}-2\bar n(1-\chi_{r})\bigr)\bigr \} -(1+\chi_{r})\bar\Omega^{2}$ 
and $\bar \beta=\Gamma_{-}\bigl\{\gamma\chi_{r}(\chi_{r}-1)-\Gamma\bigl(3+\chi_{r}+2\bar n(1+\chi_{r})\bigr)\bigr \} - (1 - \chi_{r})\bar\Omega^{2}$, 
while $\Gamma_{\pm}=\Gamma(1+2\bar n) \pm \gamma$, $\bar \Omega = \sqrt{\Gamma^{2}(1+2\bar n)^{2}+\gamma\chi_{r}(2\Gamma+\gamma\chi_{r})}$.
Depending on the initial conditions for $\langle R_{\alpha\alpha}(0)\rangle$, $\alpha \in \{e,s,a,g\}$, and the ratio $\gamma/\Gamma$ as well as environmental 
temperatures one can recover the results presented in \cite{fluor,def,deff,defff} for the same examined two-qubit system.

However in the following, we shall investigate the influence of the phonon environment on entanglement creation between the two quantum emitters in the 
presence of a coherent external source, i.e. $\Omega \not =0$. For this reason, the concurrence will be calculated in the steady-state which has been widely 
used in this sense.
\begin{figure}[t]
\includegraphics[width=8cm]{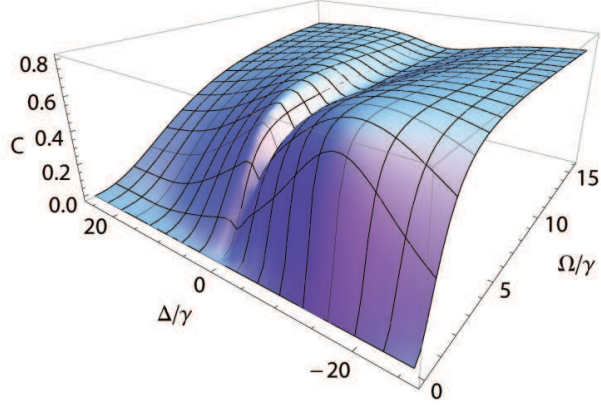}
\caption{\label{fig-1} The steady-state behavior of concurrence $C$ versus $\Delta/\gamma$ and $\Omega/\gamma$. 
The involved parameters are: $\Gamma/\gamma=3$, $\Omega_{dd}/\gamma=15$, $\chi_{r}=0.9$ and $\bar n=0.05$.}
\end{figure}

\section{Enhanced two-qubit entanglement via phonon reservoir}
For a mixed state of qubits $\{q_{1},q_{2}\}$ with density matrix $\rho_{q_{1}q_{2}}$, the concurrence $C$ is defined as \cite{entd1,entd2}
\begin{equation}
C=\rm{max}\{0,s_{1}-\sum^{4}_{\xi=2}s_{\xi}\},
\end{equation}
where the quantities $s_{\xi}$ ($\xi \in \{1,2,3,4\}$) are the square roots of the eigenvalues of the matrix product
\begin{equation}
R=\rho_{q_{1}q_{2}}(\sigma_{q_{1}y}\otimes\sigma_{q_{2}y})\rho^{\ast}_{q_{1}q_{2}}(\sigma_{q_{1}y}\otimes\sigma_{q_{2}y}),
\end{equation}
in descending order. Here, $\rho^{\ast}_{q_{1}q_{2}}$ denotes complex conjugation of $\rho_{q_{1}q_{2}}$, and $\sigma_{jy}$ are Pauli matrices for 
the two-level systems ($j \in \{q_{1},q_{2}\}$). The values of the concurrence range from zero for an unentangled state to unity for a maximally entangled 
two-particle state \cite{entd1,entd2}. The density matrix $\rho_{q_1q_2}$ can be represented in the basis $|ee\rangle$, $|eg\rangle$, $|ge\rangle$ and 
$|gg\rangle$, which is symmetric under the exchange of the sub-systems \cite{entd1,entd2,wang}. Taking into account the equations (\ref{dks}), (\ref{sep}) 
and (\ref{sec}) as well as the equations of motions involving the antisymmetrical channel $|e\rangle \leftrightarrow |a\rangle \leftrightarrow |g\rangle$, that 
is,
\begin{eqnarray}
\langle \dot R_{ea}\rangle &=& i\sqrt{2}\Omega\langle R_{sa}\rangle + i\bigl(\Delta + \Omega_{dd}+i\gamma(3-\chi_{r})/2 +i\Gamma\bar n\bigr)\nonumber \\
&\times&\langle R_{ea}\rangle, \nonumber \\
\langle \dot R_{ag}\rangle &=& i\bigl(\Delta - \Omega_{dd} + i\gamma(1-\chi_{r})/2 + i\Gamma\bar n \bigr)\langle  R_{ag}\rangle \nonumber \\
&-&i\sqrt{2}\Omega\langle R_{as}\rangle - \gamma(1-\chi_{r})\langle  R_{ea}\rangle, \nonumber \\
\langle \dot R_{sa}\rangle &=&\bigl(2i\Omega_{dd} - \gamma -\Gamma(1+2\bar n)\bigr)\langle  R_{sa}\rangle \nonumber \\
&+& i\sqrt{2}\Omega\bigl(\langle R_{ea}\rangle + \langle R_{ga}\rangle \bigr), 
\label{sea}
\end{eqnarray}
one can calculate the concurrence $C$ for $\Omega_{dd} > 0$. Again, the missing equations of motion can be obtained via Hermitian conjugation of the 
above system of three equations, i.e., Eqs.~(\ref{sea}). In fact, in the steady-state one always has: $\langle R_{ea}\rangle$=$\langle R_{ag}\rangle$=
$\langle R_{sa}\rangle=0$, and so does the corresponding Hermitian conjugate parts.
\begin{figure}[t]
\includegraphics[width=8cm]{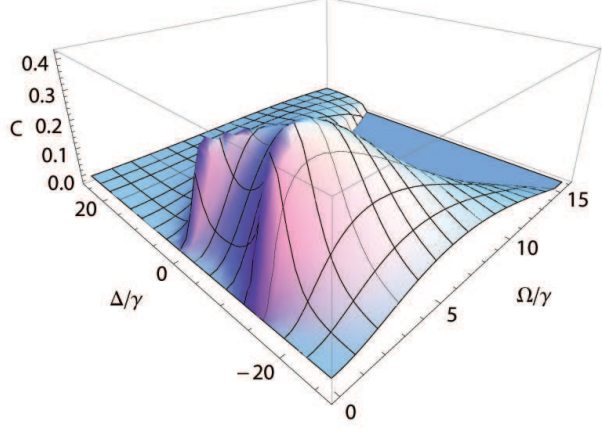}
\caption{\label{fig-2} Same as in Fig.~\ref{fig-1} but for $\Gamma/\gamma=0$, i.e., no coupling to the phonons.}
\end{figure}

\begin{figure}[t]
\includegraphics[width=4.24cm]{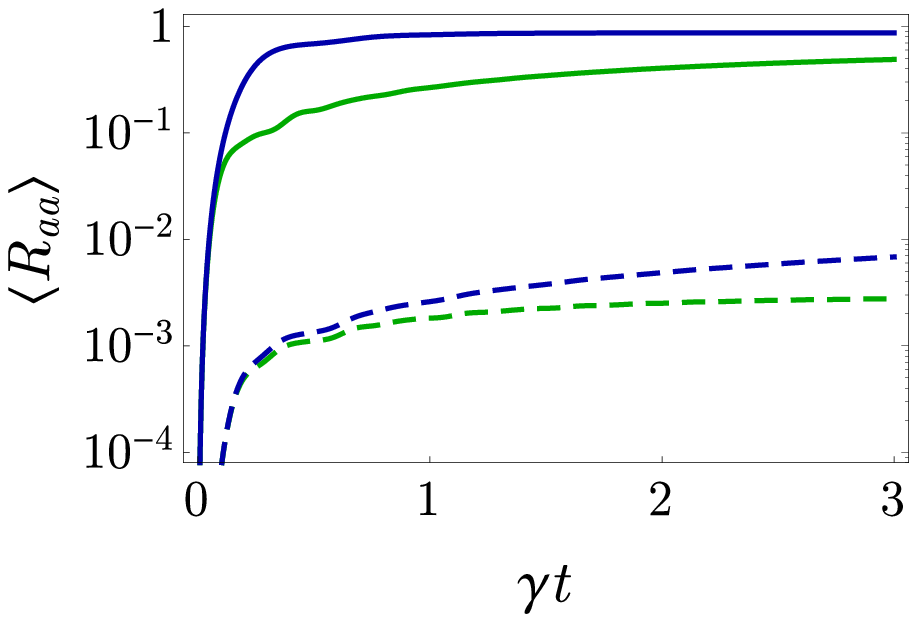}
\includegraphics[width=4.28cm]{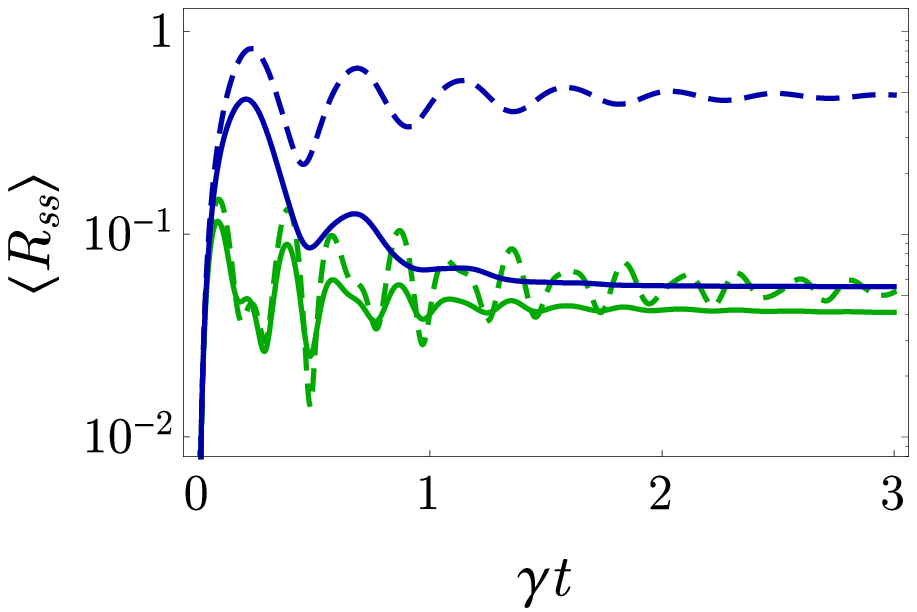}
\begin{picture}(0,0)
\put(-150,60){(a)}
\put(-20,60){(b)}
\end{picture}
\caption{\label{fig-3}The time-dependences of the populations in (a) the subradiant state $|a\rangle$ as well as in (b) the superradiant state $|s\rangle$ 
against the scaled time, when $\Omega_{dd} > 0$. The solid blue line corresponds to $\Delta = -\Omega_{dd}$ whereas the solid green curve is 
for $\Delta = \Omega_{dd}$. Respectivelly, the dashed lines stand for $\Gamma=0$. Other parameters are the same as in Fig.~\ref{fig-1} with 
$\Omega/\gamma=5$ and $\langle R_{gg}(0)\rangle =1$. 
The oscillations in (b) are due to coherent driving of the transitions $|e\rangle \leftrightarrow |s\rangle \leftrightarrow |g\rangle$.}
\end{figure}

In Figure~(\ref{fig-1}), the steady-state behaviors of the two-particle concurrence $C$ is calculated as function of the external control parameters 
$\{\Delta/\gamma,\Omega/\gamma\}$. The entanglement is substantially enhanced in comparison to the case when phonons are absent, compare 
Fig.~(\ref{fig-1}) and Fig.~(\ref{fig-2}). This happens because phonons open an additional decay channels $|s\rangle \to |a\rangle$ such that the 
antisymmetrical state $|a\rangle$ gets populated, see Figure~(\ref{fig-3}). Hence, the entanglement maximum and, correspondingly, the maximum 
population of the antisymmetrical state $|a\rangle$ is achieved when the laser is in resonance with the symmetrical two-qubit state $|s\rangle$, i.e., 
$\Delta \approx -\Omega_{dd}$, see Fig.~(\ref{fig-1}) and Fig.~(\ref{fig-2}). Actually, phonons change the phase of the cooperative two-particle 
state 
\begin{eqnarray}
|\Psi_{\phi}\rangle=\bigl(|eg\rangle + e^{i\phi}|ge\rangle\bigr)/\sqrt{2}. \label{psif}
\end{eqnarray}
Note that the antisymmetrical state $|a\rangle \equiv |\Psi_{\pi}\rangle$ is a subradiant one since it couples weakly with the vacuum modes of the 
electromagnetic field reservoir, see the dashed curves in Figure~(\ref{fig-3}). However, the phonon thermostat facilitates its population via a phonon 
induced decay and when the phonon decay rate $\Gamma$ is of the order of the single-qubit spontaneous decay rate $\gamma$, or even larger. 
Respectively, the total intensity of the spontaneously scattered photons in the steady-state, i.e., 
\begin{eqnarray}
I_{s} =\gamma\bigl\{ (1+\chi_{r})\langle R_{ss}\rangle + (1-\chi_{r})\langle R_{aa}\rangle + 2\langle R_{ee}\rangle\bigr\}, \label{ints}
\end{eqnarray}
reduces substantially in comparison to the same quantity but in the absence of phonons. Furthermore, higher bath temperatures lower the entanglement 
between the quantum dots for a fixed dipole-dipole interaction potential. This occurs because the symmetrical and the antisymmetrical two-qubit cooperative 
states tend to populate approximately equally, but weakly, for larger thermal phonon bath temperatures. More specifically, when $\gamma/\Gamma \ll 1$ 
and the laser-pumped qubits are close to each other, one has:
\begin{eqnarray}
\langle R_{aa}\rangle = \bigl(1+\bar n\bigr)/\bigl(1 + 4\bar n\bigr), \label{raa}
\end{eqnarray}
whereas
\begin{eqnarray}
\langle R_{ss}\rangle = \langle R_{ee}\rangle=\langle R_{gg}\rangle = \bar n/\bigl(1 + 4\bar n\bigr). \label{seg}
\end{eqnarray}
We observe that in this regime the populations do not depend on the external control parameters, except the temperature. If $\bar n \to 0$, 
$\langle R_{aa}\rangle \to 1$ while all other populations are almost zero. While if $\bar n \gg 1$, then $\langle R_{aa}\rangle$=$\langle R_{ee}\rangle$=
$\langle R_{ss}\rangle$ =$\langle R_{gg}\rangle$=1/4. The concurrence in this particular case, i.e., for $\gamma/\Gamma \ll 1$, is:
\begin{eqnarray}
C=\text{\rm max}\bigl\{0, (\sqrt{(1+\bar n)(1+2\bar n)}-3\bar n\bigr)/\bigl(1+4\bar n\bigr)\bigr\},
\label{concr}
\end{eqnarray}
which is zero when $\bar n=(3+\sqrt{37})/14$, while equals to unity if $\bar n =0$. For environmental temperatures up to several Kelvins and dipole-dipole 
coupling strengths of the order of $10^{12}{\rm Hz}$ one can achieve those values for $\bar n$ discussed here.

When $\Omega_{dd} < 0$, the situation is different. In this case, at lower environmental temperatures, the phonon bath opens a decay channel between 
the antsymmetrical and symmetrical two-qubit collective states, i.e., $|a\rangle \to |s\rangle$. However, the whole population is shared almost equally among 
the ground and symmetrical cooperative states, respectively, when the laser is in resonance with the latter state, that is around $\Delta \approx \Omega_{dd}$.
The complete population of the symmetrical state is avoided because the antisymmetrical state couples weakly with the vacuum modes of the electromagnetic 
field in this case and, correspondingly, less population is transferred to the state $|s\rangle$. Consequently, the entanglement is weaker in comparison to 
the case when $\Omega_{dd} > 0$, and its magnitude is close to that shown in Figure~(\ref{fig-2}).

\section{Summary}
Summarizing, we have proposed a scheme allowing to populate the subradiant two-particle collective state, which otherwise almost is decoupled from the 
interaction with the surrounding electromagnetic vacuum reservoir. The sample, consisting from an identical pair of two-level quantum dots, interacts 
coherently with a laser field and dampens via the environmental electromagnetic field reservoir as well as the phonon thermostat. The phonon bath induces 
transitions among the symmetrical and antisymmetrical two-qubit collective states, around the dipole-dipole frequency shift between them, with a subsequent 
population of the antisymmetrical cooperative state. Hence, quantum entanglement among the closely spaced constituents substantially enhances due to 
vibrational phonons and at lower temperatures.

\acknowledgments
This work was supported by a mobility project of the Romanian National Authority for Scientific Research and Innovation, CNCS - UEFISCDI, 
project number PN-III-PI-1.1-MCD-2016-0088, within PNCDI III, as well as by the Academy of Sciences of Moldova, grant No. 15.817.02.09F. 
Furthermore, we acknowledge useful discussions with Victor Ceban, Profirie Bardetski and Sergiu Cojocaru.


\end{document}